# 1-Tb/s/λ Transmission over Record 10714-km AR-HCF

**Dawei Ge[1,†], Siyuan Liu[5,1,†], Qiang Qiu[2,†], Peng Li[3,†], Qiang Guo[4], Yiqi Li[2], Dong Wang[1], Baoluo Yan[2], Mingqing Zuo[1], Lei Zhang[3], Dechao Zhang[1,*], Hu Shi[2,*], Jie Luo[3], Han Li[1], and Zhangyuan Chen[5]**

[1]*Department of Fundamental Network Technology, China Mobile Research Institute, Beijing 100053, China*
[2]*Department of Wireline Product R&D Institute, ZTE Corporation, Beijing 100029, China*
[3]*State Key Laboratory of Optical Fiber and Cable Manufacture Technology, YOFC, Wuhan, Hubei, 430074, China*
[4]*B&P Lab, Huawei Technology Co., Ltd., Shenzhen 518129, China*
[5]*State Key Laboratory of Photonics and Communications, Peking University, Beijing 100871, China*
[†]*These authors contributed equally to this work.*
**zhangdechao_cmri@sina.com*, *shi.hu@zte.com.cn*

**Abstract:** We present the first single-channel 1.001-Tb/s DP-36QAM-PCS recirculating transmission over 73 loops of 146.77-km ultra-low-loss & low-IMI DNANF-5 fiber, achieving a record transmission distance of 10,714.28 km. 

## 1. Introduction

With the massive deployment of 400G and 800G optical networks, research into next-generation beyond 1Tb/s (B1T) is actively in progressing. ITU-T SG15/Q11 has initiated the standardization of G.709.b1t for electrical interfaces supporting B1T. In single-channel beyond 1-Tb/s transmissions, the achievable transmission distance remains a critical research focus. Several significant studies utilizing solid-core single mode fiber (SMF) have been reported, as summarized in Fig. 1(a) [1-11]. These studies can be categorized based on symbol rate: 128 GBd, 135-145 GBd, 160-170 GBd, and 180 GBd. For a 1-Tb/s signal at a 128-GBd symbol rate, the maximum reported distance using G.654.E fiber and erbium-doped fiber amplifiers (EDFAs) is 1,000 km [1]. Enhancing the symbol rate to 135-145 GBd and employing low-noise-figure amplification techniques such as distributed Raman amplification (DRA) or optical parametric amplifiers (OPAs) based on periodically poled LiNbO3 (PPLN) waveguides extends the reach to 2,220 km for a 1.38-Tb/s DP-64QAM-PCS signal [3]. Further increasing the symbol rate to 168 GBd enables the use of lower entropy, pushing the transmission distance to 3,840 km [9]. However, the 10,000-km milestone, a benchmark for transoceanic submarine communications, has not yet been achieved over SMF. Recently, antiresonant hollow-core fibers (AR-HCFs) have demonstrated record-low attenuation below 0.1 dB/km [20-22], surpassing solid-core SMF. AR-HCFs offer ultra-low nonlinearity, a wide transmission window, low chromatic dispersion, and minimal latency. Especially, the ultra-low nonlinearity based on air-guiding enables much higher launch power in AR-HCF-based systems than in SMF-based systems, providing an exceptionally clean optical channel. However, prior studies on coherent transmission over HCF or hybrid HCF-SMF systems have been limited by either low data rates (e.g., 100 Gb/s) or short distances (e.g., 20 km), and have not fully exploited HCF's advantages [12-19].

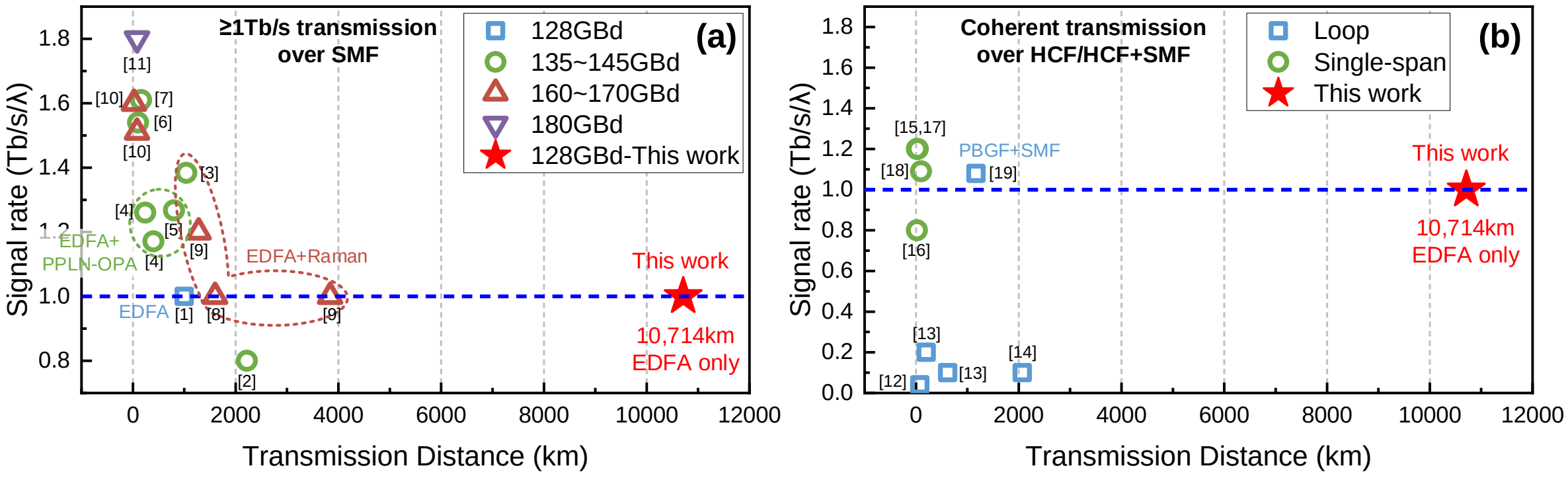

Fig. 1. Signal rate vs. distance for recent reports on (a) ≥1-Tb/s transmission over SMF, (b) Coherent transmission over HCF/HCF+SMF.

In this work, we employ a 146.77-km double-nested antiresonant nodeless fiber with five elements (DNANF-5), featuring an ultra-low overall loss of 0.134 dB/km and a low intermodal interference (IMI) of -74 dB/km. Using a single-channel DP-36QAM-PCS signal at 192.475 THz with a launch power of 23 dBm, we achieve a record transmission distance of 10,714.28 km over 73 loops of DNANF-5, with amplification provided exclusively by EDFAs. To the best of our knowledge, this represents the longest transmission distance ever reported for single-channel beyond 1-Tb/s transmission over any fiber type. IMI and attenuation ripple are also located as the dominant penalty contributors, rather than conventional nonlinearity in solid-core SMF.

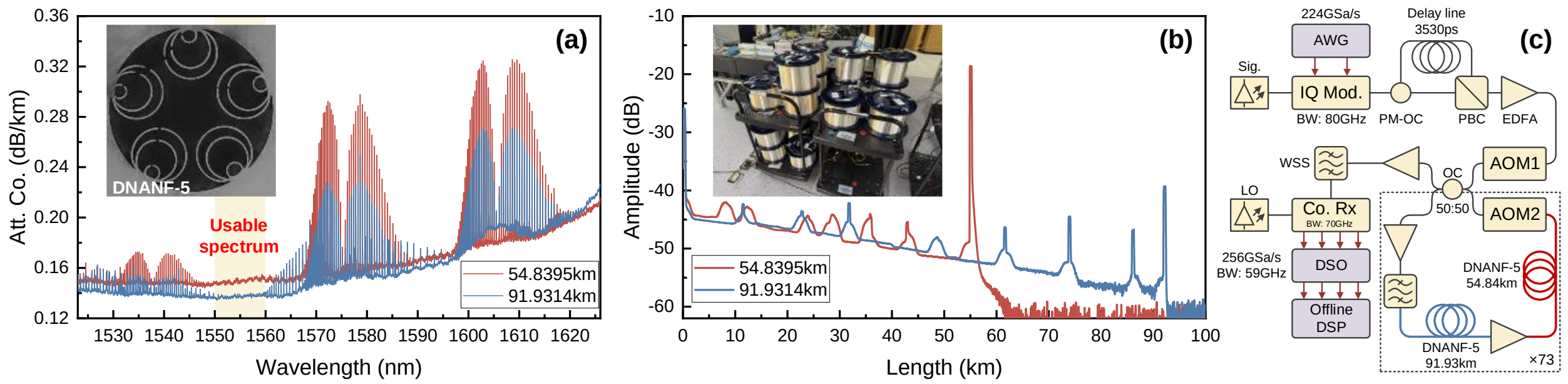

Fig. 2. (a) The attenuation spectra of the two DNANF-5 spans, inset: the SEM image of DNANF-5 used in this experiment, (b) The OTDR traces of the two DNANF-5 spans, inset: the photograph of DNANF-5 used in this experiment, (c) The schematic setup of the offline recirculating loop experiment.

## 2. Fiber Link Preparation and Recirculating System Setup

We chose to follow the relatively mature structure of DNANF-5 that has proven both ultra-low loss and low intermodal interference (IMI). A total of 311-km of DNANF-5 was fabricated, with an average attenuation of 0.15 dB/km at 1550 nm. Three primary criteria were used to select fiber spans: (1) span loss should be approximately 12 dB to ensure optimal operation of commercial EDFAs, (2) longer bobbins with lower attenuation coefficients were preferred to minimize splice-induced losses (typically 0.05-0.15 dB per splice), and (3) IMI, a key performance-limiting factor, should be as low as possible. Following these criteria, two DNANF-5 spans of 54.8395 km and 91.9314 km were selected, each comprising multiple spools (7 and 8, respectively), as shown in Fig. 2(b). HCF-SMF adapters were attached to facilitate connectivity. The attenuation spectra of the selected spans were measured using an optical spectrum analyzer (OSA, YOKOGAWA AQ6370D) with a 0.02-nm resolution, as depicted in Fig. 2(a). Notable absorption features due to water, $CO_2$, and CO narrow the usable spectrum for current ultra-long-haul transmission to approximately 10 nm (~1.24 THz). The lowest link losses for the two spans were 0.144 and 0.134 dB/km, while IMI was measured at -74 dB/km using the transmission fluctuation method [23, 24]. The chromatic dispersion (CD) coefficient ranged from 3.46 to 3.78 ps/nm/km, with a polarization mode dispersion (PMD) of 0.81 ps/km$^{1/2}$ in the C+L bands. Among the rest 164-km DNANF-5, two additional spans (30.8037km and 68.1653km) with similar attenuation but higher IMI (-64 dB/km) were also selected for comparison. It is worth noting that DNANF-5 used in this experiment was not subjected to internal gas purification, so dense wavelength division multiplexing (DWDM) with multiple channels can not be performed over such a long distance.

The schematic of the offline recirculating loop experiment is shown in Fig. 2(c). A narrow-linewidth tunable laser generated the test channel, which was modulated into a DP-36QAM-PCS signal using an 80-GHz IQ modulator (Lumentum CDM0130DC) with 128-GBd sequences from a 224-GSa/s arbitrary waveform generator (AWG, Keysight M8199B). Since the AWG is a two-port device, a polarization division multiplexing (PDM) emulator comprising a polarization-maintaining optical coupler (PM-OC), a 3530-ns delay line, and a polarization beam combiner (PBC) was used. The modulated signal was amplified by an EDFA before entering the DNANF-5 recirculating loop. Two acousto-optic modulators (AOMs) controlled the signal recirculation times and bitstream removal from the loop. The two EDFAs inside the loop operated in saturation output mode at 23 dBm. A wavelength selective switch (WSS) was employed to filter out the accumulated out-of-band noise. Each AOM pass introduced an 80 MHz frequency shift toward higher frequencies, resulting in a 6.25-GHz broadening of the in-line filtering bandwidth. At the receiver side, after amplification and filtering, the optical signal was detected using an optical modulation analyzer (Keysight N4391B-007) for offline digital signal processing (DSP), and the generalized mutual information (GMI) was used to estimate the data rate.

## 3. Experimental Results and Discussions

As shown in Fig. 3(a), DNANF-5 exhibits a “rougher” attenuation spectrum than G.652.D fiber, likely due to microbending, axis non-uniformity, or IMI. To assess the impact of the attenuation ripple, we measured the signal spectrum after different loops. As shown in Fig. 3(b), selective fading worsens with the increasing loops and transmission distance, though frequency shifts from AOMs slightly mitigate the effect. This fading likely results from DNANF-5’s attenuation ripples combined with EDFA’s gain competition. Through spectral scan, the flattest window between two major CO/$CO_2$ absorption spectra was identified near 192.475 THz for low-IMI DNANF-5 and 192.55 THz for high-IMI DNANF-5. The scan step was 25GHz, and the spectra of 10 out of these 66 channels after 70 loops (about 10,273 km) are plotted in Fig. 3(a). All these channels have completely different spectra due to attenuation ripple.

The recirculating results of both DNANF-5 links are illustrated in Fig. 3(c). The entropy of the measured signal after each loop was decided according to its normalized GMI (NGMI) and optical signal-to-noise ratio (OSNR). For low-IMI DNANF-5 (-74 dB/km IMI), 1.001 Tb/s was achieved over 10,714.28 km with 128-GBd DP-36QAM-PCS and an entropy of 3.96. In contrast, for high-IMI DNANF-5 (-64 dB/km IMI), the transmission distance of 1-Tb/s signal dropped to 3,459.8 km. OSNR measurements (Fig. 3(d)) showed an additional 10.32-dB penalty for DNANF-5 compared to back-to-back (BTB) transmission. Fig. 3(e) compares the spectra after about 3500-km transmission for the two DNANF-5 links with different IMI levels and shows that the selective fading effects are similar, indicating that the selective fading is not primarily caused by IMI but has independent origins. Thanks to the negligible nonlinearity of DNANF-5, OSNR was high enough after 10000 km transmission. Considering Fig.3(c)-(e), IMI is the dominant contributor to the transmission penalty. Nevertheless, it is still the first time that 1-Tb/s/λ signal can be transmitted over 10,000 km.

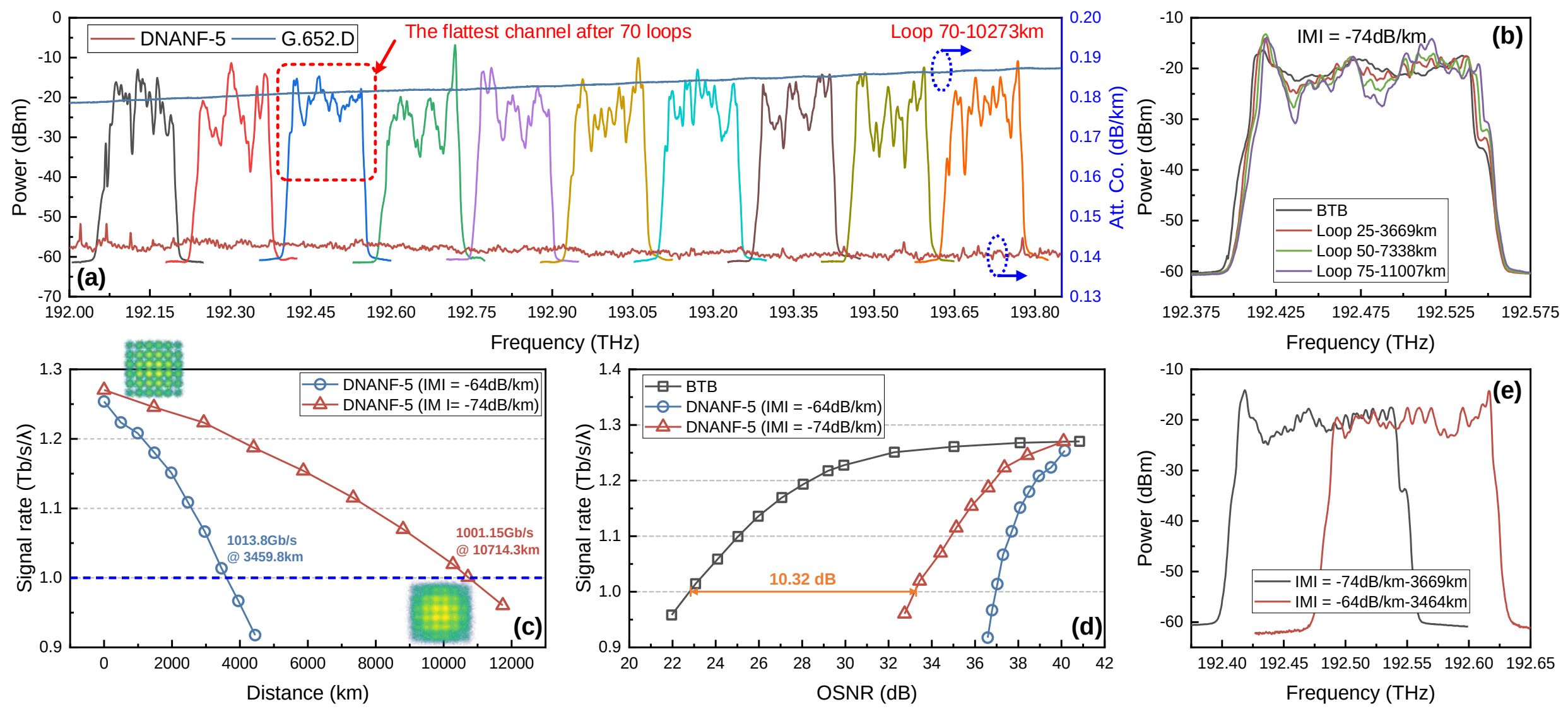

Fig. 3. (a) The spectra of different channels after 70 loops. (b) The spectra of signal after different loops for DNANF-5 link with low IMI. (c) The signal rate vs distance for different fiber links with different IMI levels. (d) The signal rate vs OSNR for DNANF-5 links and BTB. (e)The spectra of signal after about 3500km for DNANF-5 links with different IMIs.

For low-IMI DNANF-5, other channels' performances were also measured, data rates were all below 1 Tb/s, typically ranging from 400 Gb/s to 900 Gb/s due to much worse selective fading. For future applications, current AR-HCFs may already be suitable for short and middle reach scenarios, like data center interconnection (DCI) and metro backbone networks. IMI and attenuation ripple should be investigated and mitigated for long-haul transmission, and the criteria for different scenarios should be considered in future fiber standardization in ITU-T and IEC.

## 4. Conclusions

In conclusion, we successfully demonstrated the first ever 1-Tb/s s/λ signal transmission over 10,000 km without regeneration. 146.77-km low-IMI and ultra-low-loss DNANF-5 was used as the recirculating loop. A DP-36QAM-PCS with an entropy of 3.96 was modulated and transmitted after 73 loops at 23-dBm launch power. The total transmission distance reaches a record 10,714.28km. IMI and attenuation ripple take the place of nonlinearity as the dominant penalty contributors, which need further research and mitigation. *This work was supported by the Young Elite Scientists Sponsorship Program by China Association for Science and Technology (CAST) (2023QNRC001).*